\documentclass[prb,aps,superscriptaddress,floatfix,tightenlines,showpacs,twocolumn]{revtex4-2}
\usepackage{graphicx}
\usepackage{dcolumn,multirow}   
\usepackage{bm}        
\usepackage{amssymb}   
\usepackage{amsmath}
\usepackage{url}
\usepackage{layout}
\usepackage{color}
\usepackage{epsfig}
\usepackage{algorithmic}
\usepackage{booktabs}
\usepackage{float}
\usepackage{ulem}
\usepackage[hyperindex,breaklinks]{hyperref}

\definecolor{gray}{rgb}{0.5,0.5,0.5}

\begin{document}

\title{Critical behavior of Anderson transitions in higher dimensional Bogoliubov-de Gennes symmetry classes}

\author{Tong Wang}
\email{wangtong617@pku.edu.cn}
\affiliation{International Center for Quantum Materials, School of Physics, Peking University, Beijing 100871, China}
\affiliation{Collaborative Innovation Center of Quantum Matter, Beijing 100871, China}%
\author{Zhiming Pan}
\email{panzhiming@westlake.edu.cn}
\affiliation{Institute for Theoretical Sciences, Westlake University, Hangzhou 310024, Zhejiang, China}
\author{Keith Slevin}
\affiliation{Department of Physics, Osaka University, Toyonaka, Osaka 560-0043, Japan}%
\author{Tomi Ohtsuki}
\affiliation{Physics Division, Sophia University, Chiyoda-ku, Tokyo 102-8554, Japan}
\date{\today}

\begin{abstract}
Disorder is ubiquitous in solid-state systems, and its crucial influence on transport properties was revealed by the discovery of Anderson localization.
Generally speaking, all bulk states will be exponentially localized in the strong disorder limit, but whether an Anderson transition takes place depends on the dimension and symmetries of the system.
The scaling theory and symmetry classes are at the heart of the study of the Anderson transition, and the critical exponent $\nu$ characterizing the power-law divergence of localization length is of particular interest.
In contrast with the well-established lower critical dimension $d_l=2$ of the Anderson transition, the upper critical dimension $d_u$, above which the disordered system can be described by mean-field theory, remains uncertain, 
and precise numerical evaluations of the critical exponent in higher dimensions are needed.
In this study, we apply Borel-Pad\'e resummation method to the known perturbative results of the non-linear sigma model (NL$\sigma$M) to
estimate the critical exponents of the Boguliubov-de Gennes (BdG) classes.
We also report numerical simulations of class DIII in 3D, and classes C and CI in 4D, and compare the results of the resummation
method with these and previously published work.
Our results may be experimentally tested in realizations of quantum kicked rotor models in atomic-optic systems, where the critical behavior of dynamical localization in higher dimensions can be measured.
\end{abstract}

\maketitle

\section{Introduction}
Since the discovery of Anderson localization \cite{Anderson1958}, the effects of disorder in various media have been a constant focus of the physics community. 
The disorder-driven Anderson transition (AT) is a second-order quantum phase transition, around which physical observables show universal power-law behaviors. 
The universality class of the AT depends on the dimensionality and fundamental symmetries of the system: time-reversal symmetry, particle-hole symmetry, and chiral symmetry \cite{Wegner1976,Abrahams1979,EversMirlin2008}. 
Based on these symmetries, Altland and Zirnbauer (AZ) completed the symmetry classification of non-interacting disordered Hamiltonians known as the ``10-fold way" \cite{AltlandZirnbauer1997}.
The classification is comprised of the three Wigner-Dyson classes (A, AI, and AII), the three chiral classes (AIII, BDI, and CII), and the four Bogoliubov–de Gennes (BdG) classes (D, C, DIII, and CI).
The AZ classification is revelatory not only to the study of localization phenomena, but to the study of topological materials \cite{Schnyder2008,Ryu2010,chiu2016classification}.

The critical exponent $\nu$ of the AT characterizes the power-law divergence of the correlation length $\xi$ on approaching the critical point,
\begin{align}
\xi \sim |x-x_c|^{-\nu},
\label{eq:xi}
\end{align}
where $x$ is the tuning parameter and $x_c$ is the critical point. 
Constrained by computational capacity, relatively few numerical studies have gone beyond three-dimensions (3D) \cite{Garcia2008,Edge2012,Ueoka2014,Slevin2016,Prelovsek2021} 
into higher dimensions where a stronger strength of disorder is required to drive the system into localization.
A strong-disorder renormalization group (RG) approach is in development to provide theoretical insights \cite{Mard2017,Igloi2018}.
Recently, the potentials of such efforts are revealed by the proposed superuniversality of ATs in Hermitian and non-Hermitian systems \cite{Luo2022NH}, and the mapping between certain disorder-free interacting systems and disordered non-interacting systems with extra dimension \cite{Syzranov21arXiv,Syzranov23arXiv}.
Moreover, the theory and numerical simulations are applicable to experimental realizations of quantum kicked rotors with synthetic dimensions \cite{Chabe2008QKM,Lopez2012QKM,SANTHANAM20221}.

While the lower critical dimension $d_l=2$ of the AT is well established by the one-parameter scaling theory \cite{Abrahams1979}, the upper critical dimension $d_u$, above which a mean-field description is accurate, remains debatable. 
The self-consistent theory of AT by Vollhardt and W\"olfle \cite{VollhardtWolfle1980,VollhardtWolfle1982} gives the critical exponent of Anderson model (class AI) as
\begin{align}
\nu=\begin{cases}
\displaystyle \frac{1}{d-2}, & 2<d<4,	\\
\displaystyle\ \ \frac{1}{2}\ \ , & d\ge4.
\end{cases}
\label{eq:selfconsis}
\end{align}
The results that $d_u=4$ and the mean-field critical exponent $\nu=1/2$ are reminiscent of the $\phi^4$ theory.
A modified version of this theory that considers the renormalization of the diffusion coefficient \cite{Garcia2008} gives
\begin{align}\label{eq:modified}
\nu =\frac{1}{2} +\frac{1}{d-2},
\end{align}
and $d_u=\infty$. 
The prediction of the limiting value 
\begin{equation}
    \lim_{d\rightarrow \infty} \nu = \frac{1}{2},
    \label{eq:nu_limit_half}
\end{equation}
by both theories agree with the value from the Anderson model on an infinite-dimensional Bethe lattice \cite{Abou_Chacra_1973,Kunz1983Bethe,Mirlin1991,Mirlin1994,KLEIN1998163,DeLuca2014BL,Samuel2019BL}.
However, Eq. (\ref{eq:modified}) is in better agreement with numerical results\cite{Schreiber1996, Garcia2007, Slevin14, Ueoka2014} 
of the orthogonal symmetry class for $d=3,4,5,6$ than Eq.(\ref{eq:selfconsis}).

On the other hand, the nonlinear sigma model (NL$\sigma$M), an effective field theory of Anderson localization, has been studied extensively in $d=2+\epsilon$ dimensions \cite{Wegner1979,Schafer1980,Efetov1983,Hikami1992}. 
The $\beta$-function, which describes the renormalization of the conductance with system size, can be calculated analytically using perturbation techniques \cite{Hikami81,Hikami81NLsM,Hikami83,Wegner1986}. 
From the $\beta$-function one can derive relevant physical quantities including a series in powers of $\epsilon$ for the
critical exponent $\nu$.
This method, which is referred to as the $\epsilon$-expansion, is rigorous only when $\epsilon \ll 1$.
In this limit, the $\epsilon$-expansion gives $\nu = 1 / \epsilon$ in agreement with Eq. (\ref{eq:selfconsis}) but not Eq. (\ref{eq:modified})
and with numerical simulations on fractals with spectral dimensions close to $2$ \cite{Schreiber1996, Travenec2002}.
To obtain results for higher dimensions resummation methods are needed.
However, a straightforward resummation\cite{Hikami1992} of the power series for the critical exponent yields $\nu \to 0$ in
the limit $d \to \infty$ in disagreement with both Eq. (\ref{eq:selfconsis}) and Eq. (\ref{eq:modified}).
For the Wigner-Dyson classes ressumations that incorporate the correct asymptotic behaviour of the critical exponent
for $d \to \infty$ have been performed \cite{Ueoka2014,Ueoka2017} giving better agreement with numerical simulations 
\cite{Slevin14, Ueoka2014,  Slevin2016, Garcia2007} and experimental results \cite{Lemarie12010,Lopez2012QKM}.
However, a comprehensive understanding of the dimensional-dependence of the AT in different symmetry classes is still lacking.

In this paper, we focus on the BdG symmetry classes in 3D and 4D. 
The four BdG classes appear naturally in the topological superconducting system \cite{AltlandZirnbauer1997,EversMirlin2008}.
The underlying BdG Hamiltonian $H$ is invariant under the antiunitary transform of particle-hole symmetry (PHS) $\mathcal{C}=U_CK$,
\begin{align}
\mathcal{C}:\quad H \rightarrow -U_C^{\dagger} H^T U_C,
\end{align}
where $U_C$ is a unitary matrix and $K$ denotes
the operation of complex conjugation \cite{chiu2016classification}.
The BdG universality classes are realized at
the particle-hole symmetric point, $E=0$.
The particle-hole symmetry can be classified into two kinds, even ($\mathcal{C}^2=+1$) or odd ($\mathcal{C}^2=-1$).
The symmetry classes can further be characterized by time-reversal symmetry (TRS) $\mathcal{T}$.
There are four BdG classes: singlet/triplet SC (class D), singlet SC (class C), singlet/triplet SC with TRS (class DIII) and singlet SC with TRS (class CI).
Class D and class C describes BdG systems with even or odd PHS and broken TRS. 
Classes DIII and CI are characterized by a time-reversal operator $\mathcal{T}: H\rightarrow U_T H^T U_T^{-1}$, where the unitary matrix $U_T$ satisfies $U_T^2=\pm 1$. 
For classes DIII, one has PHS $\mathcal{C}^2=+1$ and TRS $\mathcal{T}^2=-1$. For class CI, one has PHS $\mathcal{C}^2=-1$ and TRS $\mathcal{T}^2=+1$.
The symmetries of the BdG classes are summarized in Table \ref{tab:BetaFunctions}.
Due to the absence of spin-rotation invariance, class D and class DIII exhibit weak antilocalization.

Below we apply the resummation method previously employed \cite{Ueoka2014,Ueoka2017} for the Wigner-Dyson symmetry classes 
to the BdG classes.
We also report simulations using the transfer matrix method for class DIII in 3D, and classes C and CI in four dimensions (4D).
We compare estimates of the critical exponent $\nu$ obtained by finite-size scaling analysis of the numerical
simulations with the results of the resummation method.
Our results show the ability of this Borel-Pad\'e analysis to give quantitative predictions of critical exponents $\nu$ for the BdG classes beyond 2D. 

The rest of the paper is organized as follows. 
In Sec. \ref{sec:BP}, we review briefly the Borel-Pad\'e resummation.
In Sec. \ref{sec:resum_nu}, we apply the Borel-Pad\'e method to the $\epsilon$-series of the critical exponent $\nu$ for the BdG classes.
In Sec. \ref{sec:resum_beta}, we apply the Borel-Pad\'e method to the $\epsilon$-series of the $\beta$-functions.
In Sec. \ref{sec:numerical} we report our numerical simulations. 
In Sec. \ref{sec:comparison} we compare the Borel-Pad\'e predictions with numerical results (both those reported here and previously published work). A summary is given in Table \ref{tab:A2}.
In Sec. \ref{sec:summary} we discuss and conclude our findings.

\begin{table*}
    \caption{\label{tab:BetaFunctions}List of the BdG symmetry classes and their transformation behavior under time-reversal, particle-hole, chiral (sublattice) (SLS) symmetries, and the presence ($\triangle$) or absence ($\times$) of SU(2) spin-rotation symmetry.
    The penultimate column shows corresponding non-compact fermionic replica non-linear sigma-model (NL$\sigma$M) manifolds.
    The last column shows the $\beta$-function\cite{Hikami81,Wegner1989,EversMirlin2008} of the four BdG symmetry classes. Here $\zeta$ is the Riemann zeta function.}
	\begin{ruledtabular}
		\begin{tabular}{c c c c c c c} 
			Class & TRS & PHS & SLS & SU(2) & NL$\sigma$M Manifold & $\beta(t)$-function \\ [0.5ex] 
			\hline
			D & $0$ & $+1$ & $0$ & $\times$ & $\mathrm{Sp}(2N)/\mathrm{U}(N)$ & $\epsilon t +t^2 -2t^3 +\displaystyle \frac{7}{2}t^4 - \frac{47}{6}  t^5 +\mathcal{O}(t^6)$  \\ 
			C & $0$ & $-1$ & $0$ & $\triangle$ & $\mathrm{O}(2N)/\mathrm{U}(N)$ &
			$\epsilon t -2t^2 -8t^3 -28t^4  -\displaystyle \frac{376}{3} t^5  +\mathcal{O}(t^6)$ \\
			DIII & $-1$ & $+1$ & $1$ & $\times$ & $\mathrm{Sp}(2N)$ & $\epsilon t +t^2 - \displaystyle  \frac{1}{2}t^3 + \frac{3}{8}t^4 -\frac{1}{8} \Big(  \frac{19}{6} +6\zeta(3) \Big) t^5 +\mathcal{O}(t^6)$ \\
			CI & $+1$ & $-1$ & $1$ & $\triangle$ & $\mathrm{O}(N)$ & $\epsilon t -2t^2 -2t^3 -3t^4 -2\Big( \displaystyle \frac{19}{6} +6\zeta(3) \Big) t^5 +\mathcal{O}(t^6)$ \\ [1ex] 
		\end{tabular}
	\end{ruledtabular}
\end{table*}

\section{Borel-Pad\'e Resummations}
\label{sec:BP}
In the scaling theory of Anderson transition \cite{Abrahams1979}, the $\beta$-function is defined as,
\begin{align}
    \beta(g) =\frac{d\ln g}{d\ln L},
\end{align}
where $g$ is the dimensionless conductance measured in units of $e^2/h$ and summed over the spins, and $L$ is the length of a $d$-dimensional cubic system. For the NL$\sigma$M description it is more convenient to work with the inverse conductance $t=1/(\pi g)$ and
\begin{align}
     \beta(t) =-\frac{dt}{d\ln L} = \frac{\beta(g)}{\pi  g}.
\end{align}
The critical point $t_c>0$ of the AT is a zero-crossing point of $\beta(t)$
\begin{align}\label{eq:tc}
\beta(t_c) =0,
\end{align}
and the critical conductance is given by $g_c = 1/(\pi t_c)$.
The critical exponent $\nu$ is related to the derivative of the $\beta$-function at the critical point,
\begin{align}
\frac{d\beta(t)}{dt}\Big|_{t=t_c} 
=-\frac{d\beta(g)}{d\ln g} \Big|_{g=g_c}
=-\frac{1}{\nu}.
\label{eq:nu_def}
\end{align}
The $\beta$-functions of the BdG classes up to the 4-loop order \cite{Hikami81,Wegner1989,EversMirlin2008} are listed in Table \ref{tab:BetaFunctions}.
Note that the coefficient of $t^5$ for class C in Table \ref{tab:BetaFunctions} differs from that given in
Table III of Ref. \cite{EversMirlin2008}.
\footnote{We use the value $-376/3$ for the coefficient of $t^5$ for class C, whereas in Table III of Ref. \cite{EversMirlin2008} it is $-376/48$. We believe the latter is a typo and that the coefficient $c_3(-2N)$ should be replaced by $16 c_3(-2N)$ so that the $\beta$-functions of classes D and C satisfy the duality relation $\beta_{\rm Sp}(t)=-2\beta_{\rm O}(-t/2)$ of the underlying NL$\sigma$M manifolds. The $\beta$-function of class D, which corresponds to the NL$\sigma$M manifold $\mathrm{Sp}(2N)/\mathrm{U}(N)$, is given in Eq. (3.7) of Ref. \cite{Wegner1989}.
We thank  Alexander D. Mirlin for private communication.}.
We also note in passing that the $\beta$-functions of the chiral symmetry classes were found to be strictly zero in all orders in perturbation theory \cite{Gade1991,Gade1993}. 

The Borel-Pad\'e resummation method is a technique for dealing with truncated and possibly divergent series. 
Given an infinite series $f$
\begin{equation}\label{eq:f_series}
   f(x) = \sum_k f_k x^k,
\end{equation}
its Borel sum is defined as
\begin{equation}
   \tilde{f}(x) = \sum_k \frac{f_k}{k!} x^k.
\end{equation}
The original series in Eq. (\ref{eq:f_series}) can be recovered by calculating the Borel transform
\begin{equation}
    f(x) = \frac{1}{x} \int_0^{\infty}  e^{-y/x} \tilde{f}(y) \mathrm{d}y. 
    \label{eq:BorelTrans}
\end{equation}
Suppose the coefficients $f_k$ are known for order $k\le l$. 
We approximate $\tilde{f}$ on the r.h.s by a rational function
\begin{equation}
    \tilde{f}(x) \approx r(x) = \frac{p(x)}{q(x)},
\end{equation}
where $p(x)$, $q(x)$ are polynomials of order $m$ and $n$, respectively,
\begin{align}
\label{eq:orders}
    p(x) = \sum_{k=0}^m p_k x^k,\qquad q(x)=\sum_{k=0}^n q_k x^k,\ q_0\equiv 1.
\end{align}
For choices of $[m,n]$ that satisfy $m+n=l$, the coefficients of the polynomials $p$ and $q$
are uniquely determined.
In some cases we require $m<n$ so that the Pad\'e approximant satisfies
\begin{equation}
    \lim_{x\rightarrow \infty} r(x) = 0.
\end{equation}
Then, the rational function $r$ can be decomposed into a sum of partial fractions
\begin{equation}
    r(x) =\sum_{j=1}^n \frac{a_j}{x-\lambda_j},
\end{equation}
where ${\lambda_j}$ are the roots of the polynomial $q(x)$. In general, the $\lambda_j$ and ${a_j}$ are complex numbers.
Substituting the above equation into Eq. (\ref{eq:BorelTrans}) and performing the integration, 
we obtain the Borel-Pad\'e approximation $F$ of the series for $f$
\begin{equation}
    F(x) = \frac{1}{x} \sum_{j=1}^n a_j B\left(\frac{\lambda_j}{x}\right).
\end{equation}
Here, the function $B$ is defined by
\begin{align}
    B(s) =\begin{cases}
    -\exp(-s) \mathrm{E_i}(s) \quad & s\in \mathbb{R}, s\neq 0,  \\
    \exp(-s) \mathrm{E_1}(-s) \quad & s\in \mathbb{C}, \arg s\neq \pi,
    \end{cases}
\end{align}
where
\begin{align}
\mathrm{E_i}(x) =& - \int_{-x}^{\infty} \frac{e^{-t}}{t}dt = \int_{-\infty}^{+x} \frac{e^t}{t} dt, \notag\\
\mathrm{E_1}(z) =& \int_{z}^{\infty} \frac{e^{-t}}{t}dt,\quad |\arg z|<\pi .
\end{align}

\section{Resummation of the series for $\nu ( \epsilon )$}
\label{sec:resum_nu}

Series in powers of $\epsilon$ for the critical exponent $\nu$ can be derived starting from
the series for the $\beta$-function in powers of $t$ as follows.
We take symmetry class C as an example.
We first find an approximation for $t_c(\epsilon)$ by solving Eq. (\ref{eq:tc}) using the available terms in
the power series for $\beta(t)$. 
For class C we find
\begin{equation}
    t_c(\epsilon) = \frac{1}{2}\epsilon - \epsilon^2 + \frac{9}{4}\epsilon^3 - \frac{77}{12}\epsilon^4 + \mathcal{O}(\epsilon^5).
\end{equation}
Here we have chosen the root for which
\begin{equation}
    \lim_{\epsilon \rightarrow 0} t_c = 0.
\end{equation}
If we then substitute the series for $t_c$ into Eq. (\ref{eq:nu_def}), we obtain the following series in powers of $\epsilon$
for the inverse of $\nu$
\begin{equation}
   \frac{1}{\nu}  \left( \epsilon \right) = \epsilon +2\epsilon^2 - \epsilon^3 + \frac{15}{2}\epsilon^4 + \mathcal{O}(\epsilon^5).
\end{equation}
Taking the reciprocal of this series we obtain
\begin{equation}
    \nu(\epsilon) = \frac{1}{\epsilon} -2 + 5 \epsilon -\frac{39}{2}\epsilon^2 +\mathcal{O}(\epsilon^3).
\end{equation}
Similarly, for symmetry class CI we find
\begin{align}
    t_c(\epsilon) &= \frac{1}{2}\epsilon -\frac{1}{4}\epsilon^2 + \frac{1}{16}\epsilon^3 - \frac{1+9\zeta(3)}{24} \epsilon^4 + \mathcal{O}(\epsilon^5) \nonumber \\
   \frac{1}{\nu}  \left( \epsilon \right) &= \epsilon +\frac{1}{2}\epsilon^2 +\frac{1}{4}\epsilon^3 + \frac{5+36\zeta(3)}{16}\epsilon^4 + \mathcal{O}(\epsilon^5) \nonumber \\
    \nu(\epsilon) &=\displaystyle \frac{1}{\epsilon} - \frac{1}{2} - \frac{3+36\zeta(3)}{16}\epsilon^2 +\mathcal{O}(\epsilon^3).
\end{align}
This approach works for symmetry classes C and CI because the coefficient of the $t^2$ term in $\beta(t)$ is negative
and the lower critical dimensions for these classes is $d_l=2$. 
However, for symmetry classes D and DIII the coefficient of the $t^2$ term in $\beta(t)$ is positive, so that
when we follow the procedure explained above we find
\begin{equation}
    \lim_{\epsilon \rightarrow 0} t_c \neq 0,
\end{equation}
and we are unable to obtain a useful series in powers of $\epsilon$ for $\nu$.
This reflects the possibility that the lower critical dimensions for these two classes is below 2D ($d_l<2$), as thought to be the case for the symplectic class AII.

Now we apply the Borel-Pad\'e resummation introduced in the previous section. 
A naive resummation tacitly assumes the limiting behavior 
\begin{equation}
    \lim_{d\rightarrow \infty} \nu = 0,
    \label{eq:nu_limit_zero}
\end{equation}
which disagrees with self-consistent theories of the AT and the 
results for the AT on the Bethe lattice, i.e., with Eq. (\ref{eq:nu_limit_half}).
Instead, we rewrite
\begin{equation}
    \nu \left( \epsilon \right) = \frac{1}{2} + \frac{1}{\epsilon} f \left( \epsilon \right),
\end{equation}
and perform the resummation of $f(\epsilon)$ with the requirement $m \le n$. 
Such a treatment guarantees the limiting behavior given in Eq. (\ref{eq:nu_limit_half}).
Of course, the application of this restraint to the BdG symmetry classes needs to be justified.
For later reference, in Table \ref{tab:BorelPadeNu}, we compare the results given by imposing Eq. (\ref{eq:nu_limit_half}) and Eq. (\ref{eq:nu_limit_zero}) for the classes C and CI in 3D and 4D.

\begin{table}

	\begin{ruledtabular}
		\caption{\label{tab:BorelPadeNu}Comparison of the critical exponents $\nu$ for classes C and CI in 3D and 4D 
                obtained from Borel-Pad\'e resummations of the series for $\nu(\epsilon)$ when imposing different limiting
                conditions, i.e., Eq.~(\ref{eq:nu_limit_zero}) compared with
                Eq.~(\ref{eq:nu_limit_half}).
                Numbers in the square brackets indicate the orders
                of polynomials, $m$ and $n$ [Eq.~(\ref{eq:orders})].}
		\centering
		\begin{tabular}{ccccc}
			(a) 3D & \multicolumn{2}{c}{$\displaystyle \lim_{d\rightarrow\infty} \nu = 0$} & \multicolumn{2}{c}{$\displaystyle \lim_{d\rightarrow\infty} \nu = \frac{1}{2}$} \\
			class & $[0,3]$ & $[1,2]$ & $[0,3] $ & $[1,2]$ \\  [0.5ex] 
			\hline
			C & 0.357 & 0.227 & 0.773 & 0.360 \\
			CI & 0.555 & 0.776 & 0.924 & 1.226
			\\
			\\
			(b) 4D & \multicolumn{2}{c}{$\displaystyle \lim_{d\rightarrow\infty} \nu = 0$} & \multicolumn{2}{c}{$\displaystyle \lim_{d\rightarrow\infty} \nu = \frac{1}{2}$} \\ 
			class & $[0,3]$ & $[1,2]$ & $[0,3] $ & $[1,2]$ \\  [0.5ex] 
			\hline
			C & 0.111 & -0.050 & 0.580 & 0.527 \\
			CI & 0.185 & 0.329 & 0.633 & 0.876
		\end{tabular}
	\end{ruledtabular}
\end{table}

\section{Resummation of the series for $\beta(t)$}
\label{sec:resum_beta}
\begin{figure}[t]
\centering
\includegraphics[width=0.6\linewidth]{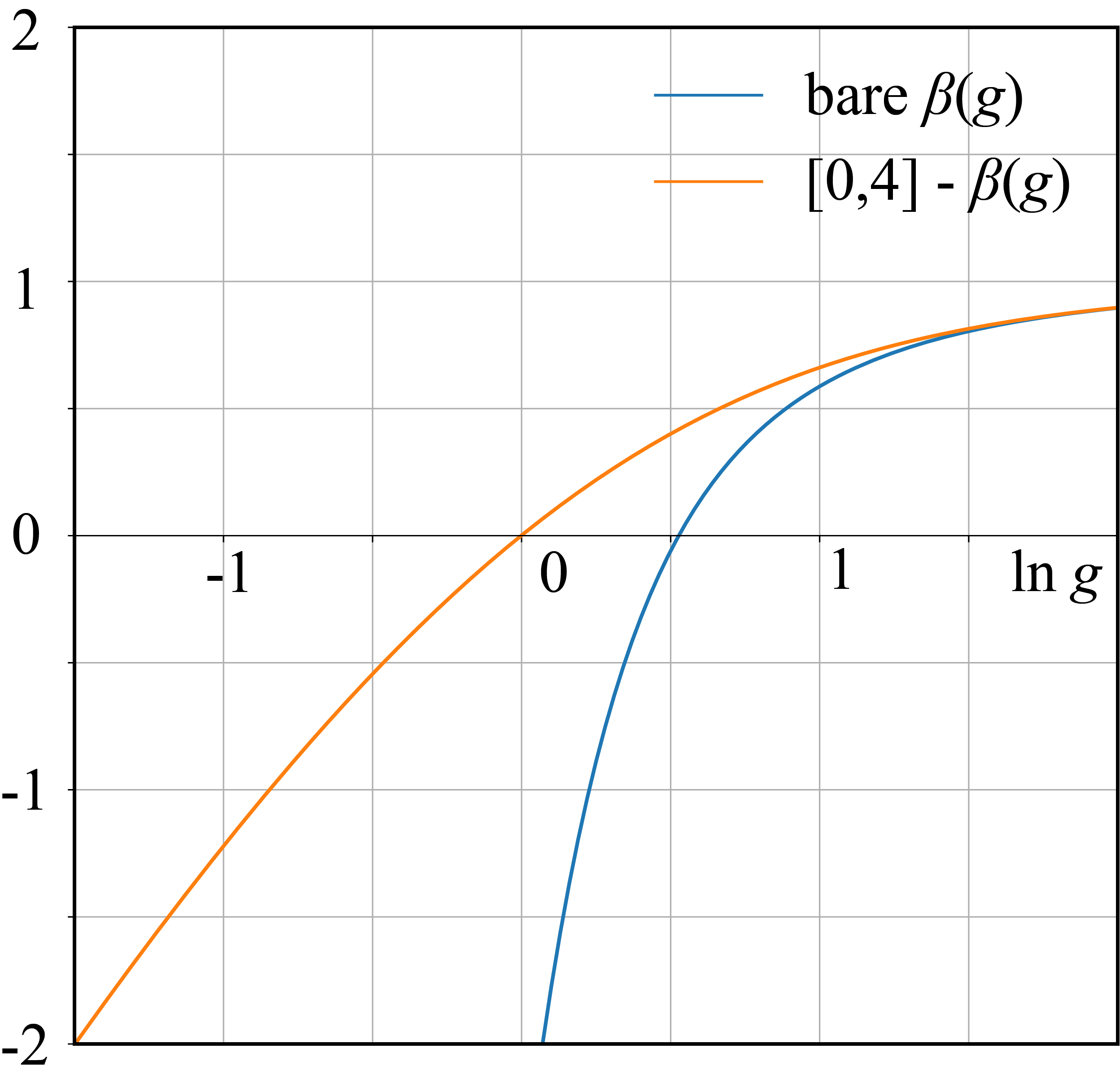}
\caption{Comparison of the approximations for the $\beta(g)$-function before and after Borel-Pad\'e resummation
of the series for class C in 3D.
Numbers in the square brackets indicate the orders
of polynomials, $m$ and $n$ [Eq.~(\ref{eq:orders})].}
\label{fig:ClassC_plot}
\end{figure}

\begin{figure}[t]
\centering
\includegraphics[width=0.6\linewidth]{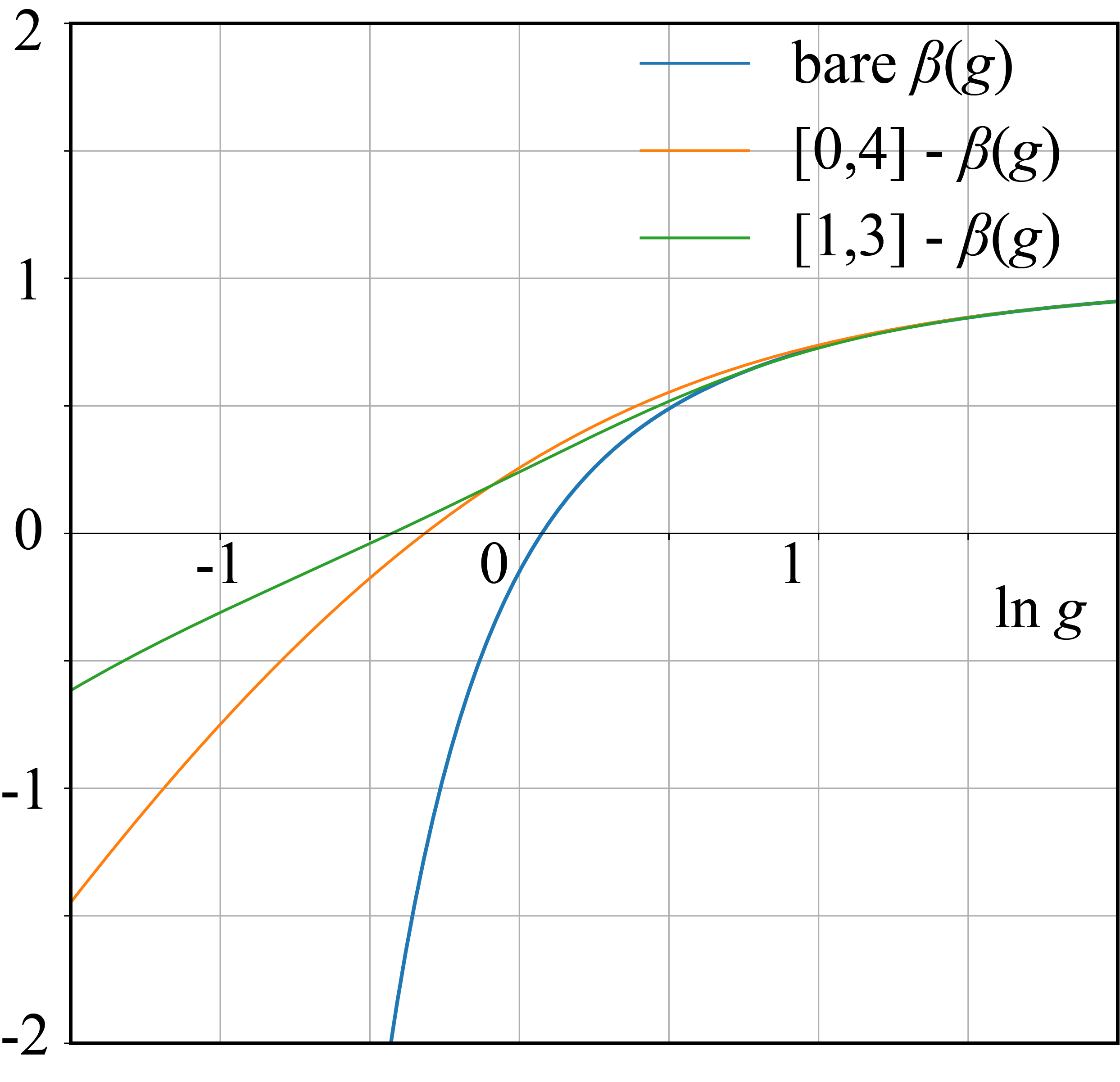}
\caption{Comparison of the approximations for the $\beta(g)$-function before and after Borel-Pad\'e resummation
of the series for class CI in 3D.}
\label{fig:ClassCI_plot}
\end{figure}

An alternative to the approach above is to apply the Borel-Pad\'e method directly to the series for the $\beta$-function.\cite{Ueoka2014}.
All the series take the form
\begin{equation}
\beta(t) = \epsilon t - t f(t),
\end{equation}
where $f$ is a power series in $t$.
In terms of $f(t)$ the critical exponent is
\begin{equation}
\frac{1}{\nu}
= t\frac{df(t)}{dt} \Big|_{t=t_c}.
\end{equation}
We need to impose the limiting behaviour at infinite dimension given in Eq.(\ref{eq:nu_limit_half}).
We first note that in high dimensions the Anderson transition takes place at strong disorder and, moreover, that
\begin{equation}
    \lim_{d\rightarrow\infty} t_c = \infty
\end{equation}
This means that we can obtain the correct limiting behaviour by arranging that
\begin{equation}\label{eq:lim_tdfdt}
    \lim_{t\rightarrow\infty} t\frac{df}{dt} = A,
\end{equation}
with $A=2$.
To do so, we define $h$, a polynomial in $t$, by
\begin{equation}\label{eq:h(t)}
	h(t) = t\frac{df(t)}{dt} - A .
\end{equation}
Applying the Borel-Pad\'e method to $h$, we obtain an approximation $H$ for $h$
that satisfies
\begin{equation}
    \lim_{t\rightarrow\infty} H(t) = 0,
\end{equation}
so that Eq. (\ref{eq:lim_tdfdt}) is satisfied.
To obtain the corresponding approximation $F$ for $f$, a further integration is needed,
\begin{align}
f(t) \approx F(t) =\int_{0}^{t} \frac{A+H(t)}{t} dt .
\end{align}
The result can be expressed in the form\cite{Ueoka2014}
\begin{align}
F(t) = \sum_{j=1}^{n} c_j B(\lambda_j/t),\quad 
c_j =\frac{a_j}{\lambda_j} .
\end{align}
Finally, the $\beta$-function is approximated as
\begin{align}
\beta(t) \approx \epsilon t - t F(t).
\end{align}
We show the resulting Borel-Pad\'e approximations for $\beta(g)$ in 3D for classes C, CI in 
Fig.~\ref{fig:ClassC_plot} and Fig.~\ref{fig:ClassCI_plot}, respectively, together with the series without resummation. 
We omit the $[m,n]=[1,3]$ resummation for class C because the resulting $\beta$-function is not monotonic and has two unphysical fixed points. The limiting behavior $\beta(g) \sim 2\ln g$ at $g\ll 1$ guaranteed by the constraint $A=2$ in Eq. (\ref{eq:lim_tdfdt}) is observed only at $\ln g$ much smaller than the range plotted in Fig.~\ref{fig:ClassC_plot} .

\begin{figure}[t]
\centering
\includegraphics[width=0.6\linewidth]{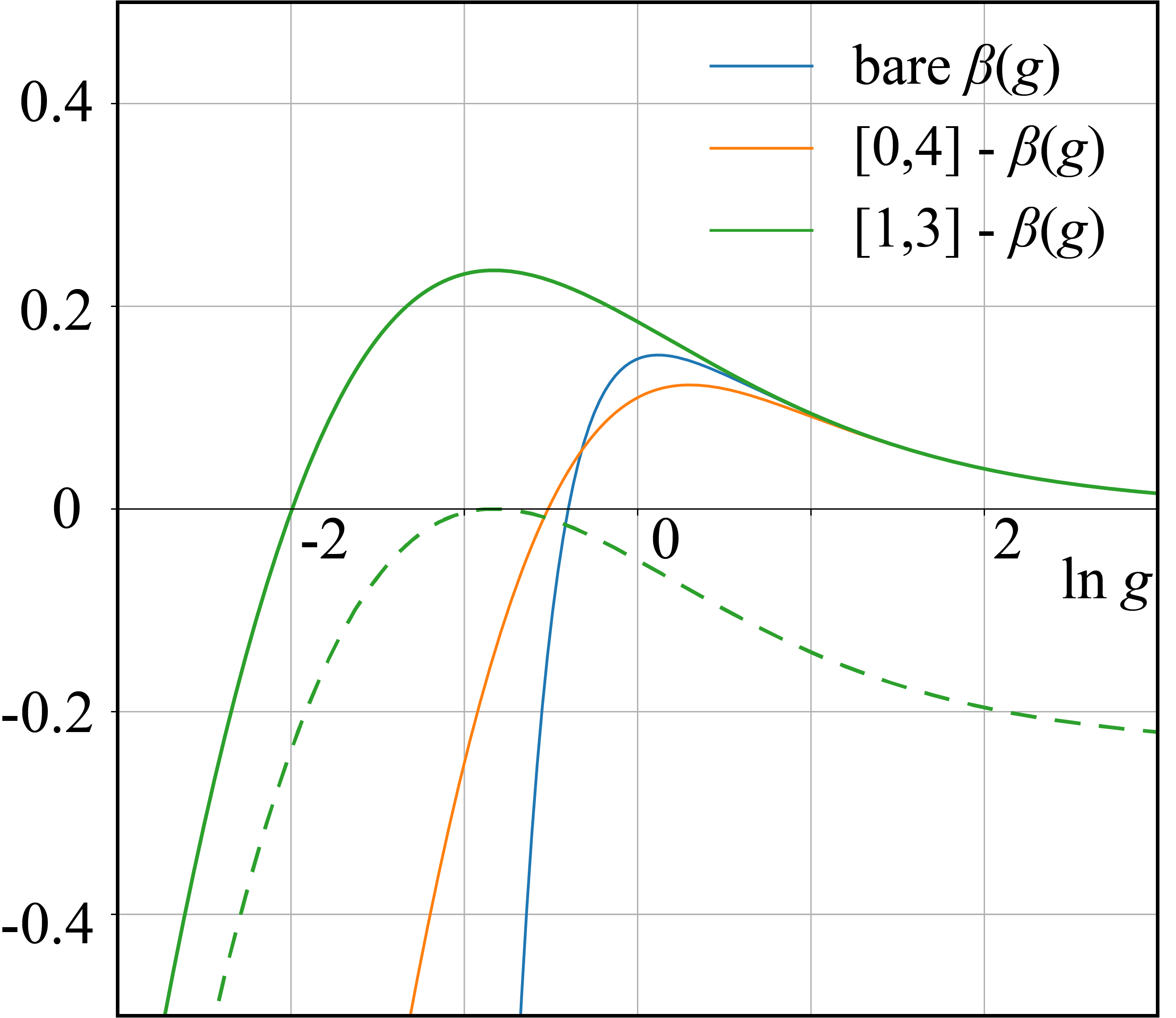}
\caption{Comparison of the approximations for the $\beta(g)$-function before and after Borel-Pad\'e resummation
of the series for class D in 2D.
The $[1,3]$ Bore-Pad\'e resummation of the $\beta(g)$-function at the corresponding estimate $d_l=1.76$ of the lower critical 
dimension is plotted with a dashed line.}
\label{fig:ClassD_plot}
\end{figure}

\begin{figure}[t]
\centering
\includegraphics[width=0.6\linewidth]{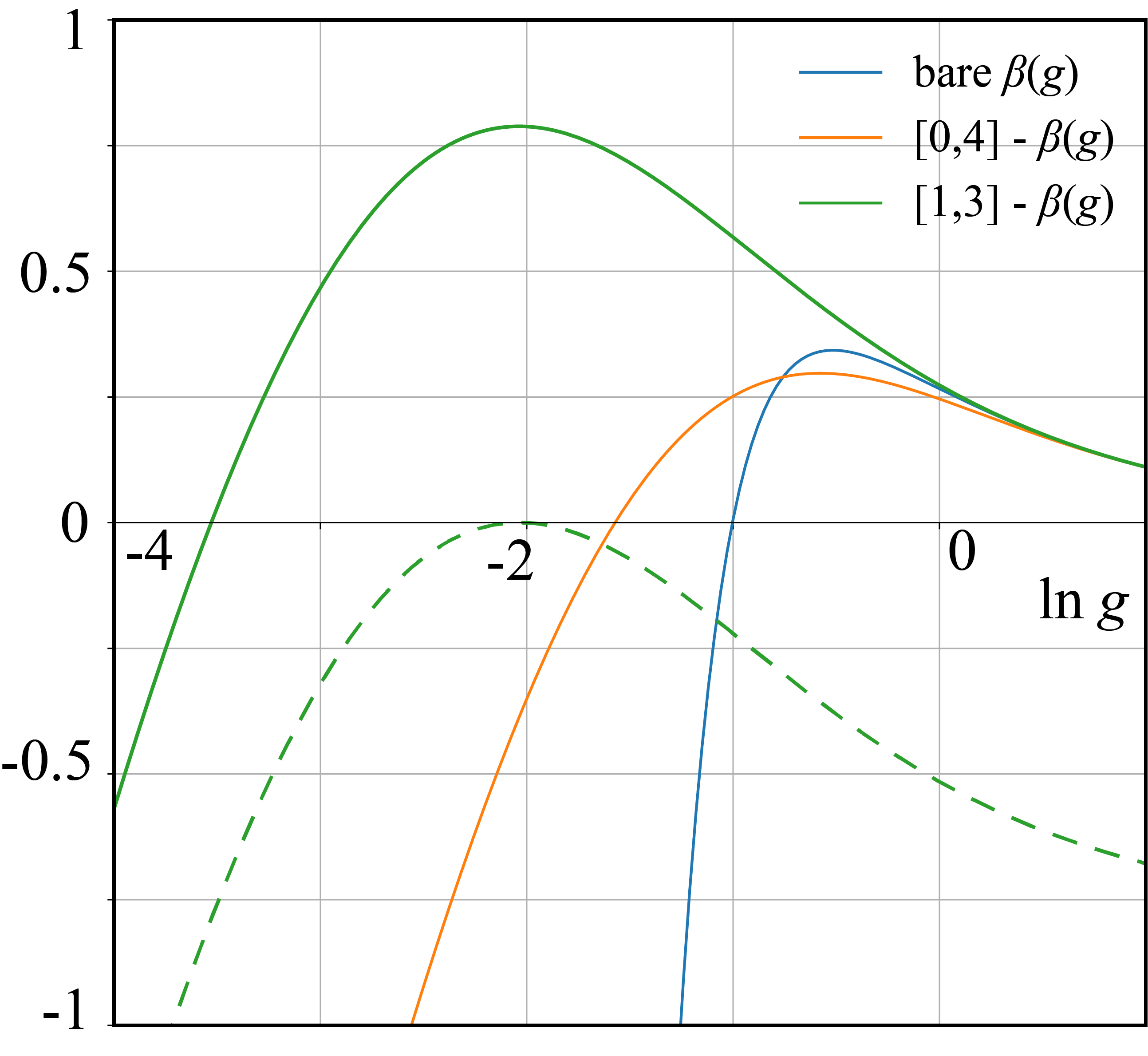}
\caption{Comparison of the approximations for the $\beta(g)$-function before and after Borel-Pad\'e resummation
of the series for class DIII in 2D.
The $[1,3]$ Bore-Pad\'e resummation of the $\beta(g)$-function at the corresponding estimate $d_l=1.21$ of the lower critical 
dimension is plotted with a dashed line.}
\label{fig:ClassDIII_plot}
\end{figure}

We show the resulting Borel-Pad\'e approximations for $\beta(g)$ in 2D for classes D and DIII in 
Fig.~\ref{fig:ClassD_plot} and Fig.~\ref{fig:ClassDIII_plot}, respectively, together with the series without resummation.
In classes D and DIII, for $d<2$, two fixed points appear: a critical fixed point, and a stable fixed point. 
At the lower critical dimension $d_l$, these two fixed points annihilate, e.g., the dashed curve in Fig.~\ref{fig:ClassD_plot} and Fig.~\ref{fig:ClassDIII_plot}, 
and the value of the $\beta$-function at its maximum is zero
\begin{align}
\max_{d=d_l} \beta(g)=0.
\end{align}
This leads directly to an estimate for $d_l$,
\begin{align}
d_l\approx 2-\max \beta(g,\epsilon=0).
\end{align}
Estimates of the lower critical dimension obtained from the Borel-Pad\'e resummations
are summarized in Table \ref{tab:dc1lcd}.

\begin{table}
\begin{ruledtabular}
    \centering
    \begin{tabular}{cclc}
     class & no resummation & \multicolumn{2}{c}{Borel-Pad\'e}  \\
      & &  $[0,4]$ & $[1,3]$    \\
     \hline
     D & $1.85$ & $1.88$ & $1.76$ \\
     DIII & $1.66$ & $1.70$ & $1.21$  \\
    \end{tabular}
    \caption{\label{tab:dc1lcd}Lower critical dimension of the BdG symmetry classes D and DIII obtained from the $\beta$-function without resummation and with order $[m,n]$ Borel-Pad\'e resummation.}
\end{ruledtabular}
\end{table}

\section{Numerical Simulations}
\label{sec:numerical}
To evaluate the effectiveness of the Borel-Pad\'e resummation in estimating the critical exponents of the BdG symmetry classes,
especially in high spatial dimensions $d\ge 3$, we perform simulations for 3D class DIII, 4D class C, and 4D class CI.
We set the energy $E$ to the particle-hole symmetric
point, $E=0$, and vary the disorder strength $W$.

\subsection{3D class DIII}
\label{subsec:3D}
This symmetry class describes time-reversal symmetric superconductors with broken spin-rotational symmetry. 
We study a four-band tight-binding model on cubic lattice \cite{liu2010model,Kobayashi2013},
\begin{align}
    \mathcal{H}_{\rm DIII}& = \sum_{\mathbf{r},\mathbf{r'}}  c_{\mathbf{r}}^{\dagger} [H_{\rm DIII}]_{\mathbf{r}\mathbf{r}'} c_{\mathbf{r}'} \nonumber \\
    & =\sum_{\mathbf{r}} \sum_{\mu=1}^3\left[\frac{i t}{2} c_{\mathbf{r}+\mathbf{e}_{\mu}}^{\dagger} \alpha_{\mu} c_{\mathbf{r}}-\frac{m_{2}}{2} c_{\mathbf{r}+\mathbf{e}_{\mu}}^{\dagger} \beta c_{\mathbf{r}}+\text { H.c. }\right] \nonumber \\
    &+\sum_{\mathbf{r}}\left(m_{0}+3 m_{2}+v_{\mathbf{r}} \right)  c_{\mathbf{r}}^{\dagger} \beta c_{\mathbf{r}}
    \label{eq:DIII}
\end{align}
where $c^{\dagger}_{\mathbf{r}}$ ($c_{\mathbf{r}})$ is the 4-component creation (annihilation) operator on a cubic-lattice site $\mathbf{r}$. For convenience we set the lattice constant $a$ to be unity. The $\mathbf{e}_{\mu=1,2,3}$ are the primitive lattice vectors along the $x,y,z$ directions, respectively. The matrices $\alpha_{\mu}$ and $\beta$ are defined as
\begin{align}
    \alpha_{\mu} = \left(\begin{array}{cc} 
    0 & \sigma_{\mu} \\ \sigma_{\mu} & 0 \end{array}\right), \quad \beta = \left(\begin{array}{cc}
    1 & 0 \\ 0 & -1 \end{array}\right),
\end{align}
where $\sigma_{\mu}$ and $\tau_{\mu}$ are Pauli matrices acting on different degrees of freedom (e.g., spin and orbital). Parameter $m_0$ is a mass, and parameters $m_2$ and $t$ are hopping amplitudes.
This Hamiltonian has time-reversal symmetry $U_T^{\dagger}\;H_{\rm DIII}^{*}\; U_T =  H_{\rm DIII}$ where
\begin{align}
	U_T = \delta_{\mathbf{rr}'} (\sigma_2 \otimes \tau_0) , \quad U_T^T=-U_T,
\end{align}
and a particle-hole symmetry $U_S^{\dagger}\; H_{\rm DIII}\; U_S = -H_{\rm DIII}$ where
\begin{align}
	U_S = \delta_{\mathbf{rr}'} (\tau_0 \otimes \tau_2).
\end{align}
This model depicts a 3D $\mathbb{Z}$ topological
insulator (TI) when $m_0 < 0$ and a trivial insulator when $m_0>0$. 

\begin{figure}
    \centering
    \includegraphics[width=\linewidth]{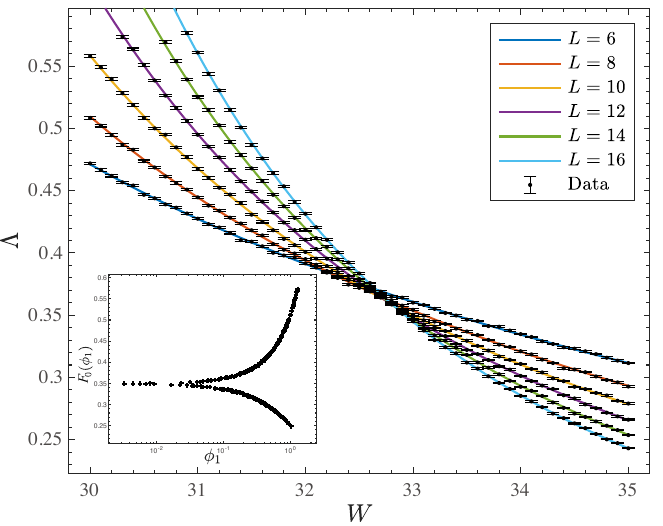}
    \caption{the dimensionless ratio $\Lambda$ near the Anderson transition for the 3D class DIII model. The expansion order is $(n_1,n_2) =(3,1)$ $(m_1,m_2)=(2,0)$ as defined in Eqs.~(\ref{eq:n1n2}, \ref{eq:m1m2}). The solid lines are the fitting functions, and the black dots with error bars are simulation data points. Inset: the scaling function }
    \label{fig:3dDIII}
\end{figure}

For numerical calculations, we specify the parameters $t=2$, $m_2=1$, $m_0=-2.5$, and use independent uniform distributions for the random on-site potential
\begin{align}
	v_{\mathbf{r}}\in [-W/2,W/2], \quad \langle v_{\mathbf{r}} v_{\mathbf{r}'} \rangle =\delta_{\mathbf{rr}'} W^2/12.
	\label{eq:vr}
\end{align}
Here, $\langle \cdots \rangle$ indicates a disorder average.
We use the transfer matrix method to calculate the localization length of the model \cite{Slevin14} and impose periodic boundary
conditions in the transverse direction.
We simulate a semi-infinite bar with a cross section of size $L\times L$ and estimate the quasi-one-dimensional (Q1D) localization length $\lambda$ at disorder strength $W$ and linear size $L$. A dimensionless ratio $\Lambda$ is defined as
\begin{align}
	\Lambda(W,L) = \lambda(W,L)/L.
\end{align}
The results are shown in Fig.~\ref{fig:3dDIII} where $\Lambda$ is plotted versus $W$ for various $L$. 
Curves for different $L$ have an approximate common crossing point.
This point indicates the Anderson transition between the TI (localized) phase and the metallic (extended) phase.

To estimate the critical exponent, we fit the data to the following scaling form that includes corrections
to single parameter scaling due to an irrelevant scaling variable \cite{Slevin99,Slevin14}
\begin{align}
    \Lambda = F\big( \phi_1, \phi_2 \big) = F\big( u_1(w) L^{1/\nu}, u_2(w) L^{-y} \big),
    \label{eq:scaling}
\end{align}
where
\begin{align}
	\omega = (W-W_c)/W_c,
\end{align}
and $\phi_1=u_1L^{1/\nu}$ is the relevant scaling variable that encodes the power-law divergence of correlation length $\xi\sim |u_1(w)|^{-\nu}$ around the critical point. The second scaling variable $\phi_2=u_2 L^{-y}$ with exponent $-y<0$ is the leading irrelevant correction, and vanishes in the limit $L\to\infty$.
We approximate the scaling function $F$ using a truncated Taylor series near the critical point ($|w|\ll 1$),
\begin{align}
    F\left( \phi_1, \phi_2 \right) &= \sum_{j=0}^{n_2} F_j(\phi_1) \phi_2^j =\sum_{i=0}^{n_1} \sum_{j=0}^{n_2} f_{ij} \phi_1^i \phi_2^j,
    \label{eq:n1n2}
\end{align}
and
\begin{align}
    u_1 = \sum_{k=1}^{m_1} b_{k} w^k,\quad u_2 = \sum_{k=0}^{m_2} c_{k} w^k.
    \label{eq:m1m2}
\end{align}
We set $b_1=c_0=1$ to remove the arbitrariness of the expansion coefficients. 
The numerical data are fitted to the scaling function by minimizing the $\chi$-squared statistic
\begin{align}
    \chi^2 = \sum_{n=1}^{N_{\rm D}} \frac{(\Lambda_n - F_n)^2}{\sigma_n^2}.
    \label{eq:chi2}
\end{align}
Here, $N_{\rm D}$ is the number of data points, $\Lambda_n$ is the value of $\Lambda$ for $n$th data point, $\sigma_n$ its standard error, and $F_n$ the value of the scaling function for the $n$th data point.
To assess whether or not the fit is acceptable, we use the goodness of fit probability.
Here, this is well approximated by \cite{Slevin14}
\begin{align}
    \mathrm{GoF} \approx 1 - \frac{1}{\Gamma(N_{\rm F}/2)}\int_0^{\chi_{\rm min}^2/2} \mathrm{d}t\, e^{-t}\, t^{\chi_{\rm min}^2/2-1},
    \label{eq:GOF}
\end{align}
where $N_{\mathrm{F}} = N_{\mathrm{D}} - N_{\mathrm{P}}$ is the degrees of freedom (with $N_{\mathrm{P}}$ the number of fitting parameters),
$\chi_{\mathrm{min}}^2$ is the minimum value of the $\chi$-squared statistic,
and $\Gamma$ is the Gamma function. 
The fitting results are shown in Table \ref{tab:fit} (a).
Our estimate of the critical exponent for 3D Class DIII is
\begin{align}
\label{eq:nuDIII}
   \nu = 0.96 \pm 0.01
\end{align}

\begin{table*}
        \caption{\label{tab:fit}FSS results for class DIII in 3D,  and classes C and CI in 4D. The orders of the expansion of the scaling function are fixed at $n_1=3$ and $n_2=1$. Here, $m_1$ and $m_2$ and the orders, respectively, of the expansions of $u_1$ and $u_2$ (see Eq.~(\ref{eq:n1n2},\ref{eq:m1m2}). The values enclosed in square brackets are 95\% confidence intervals determined from 1000 Monte Carlo samples.}
	\flushleft
	(a) 3D class DIII
	\begin{ruledtabular}
		\begin{tabular}{cccccccc}
        $L$ & $m_1$ &$m_2$&  GoF & $ W_c$ & $\nu$ & $y$ & $\Lambda_c$ \\
        \hline
        \multirow{2}*{4-16} & 2 & 0 & 0.19 & 32.909 [32.882, 32.935] & 0.972 [0.958, 0.986] & 2.09 [1.94, 2.25] & 0.349 [0.347, 0.351] \\
         & 3 & 0& 0.42 & 32.903 [32.877, 32.933] & 0.981 [0.966, 0.994] & 2.14  [1.95, 2.30] & 0.350 [0.347, 0.352] \\
        \multirow{2}*{6-16} & 2 & 0 & 0.50 & 32.917 [22.642, 22.727] & 0.952 [0.928, 0.974] & 1.98 [1.62, 2.50] & 0.349 [0.345, 0.352] \\
         & 3 & 0 & 0.60 & 32.898 [32.854, 32.965] & 0.963 [0.917, 0.979] & 2.23 [1.50, 2.94] & 0.351 [0.343, 0.354] \\
        \end{tabular}
	\end{ruledtabular}

    \flushleft
    (b) 4D class C
 	\begin{ruledtabular}
		\begin{tabular}{cccccccc}
        $L$ & $m_1$ & $m_2$ & GoF & $ W_c$ & $\nu$ & $y$ & $g_c$ \\
        \hline
        \multirow{2}*{4-12} & 2 & 0 & 0.40 & 22.65 [22.62, 22.69] & 0.724 [0.698, 0.750] & 1.45 [1.26, 1.69] & 0.83 [0.78, 0.89] \\
         & 3 & 0 & 0.49 & 22.66 [22.62, 22.70] & 0.724 [0.699, 0.751]& 1.45 [1.27, 1.71] & 0.83 [0.78, 0.89] \\
        \multirow{2}*{6-12} & 2 & 0 & 0.44 & 22.68 [22.64, 22.73] & 0.698 [0.649, 0.734] & 1.66 [1.22, 2.46] & 0.80 [0.74, 0.85] \\
         & 3 & 0 & 0.48 & 22.68 [22.64, 22.72] & 0.703 [0.652, 0.742] & 1.61 [1.18, 2.29] & 0.80 [0.75, 0.86] \\
    	\end{tabular}
	\end{ruledtabular}
	
	\flushleft
	(c) 4D class CI
        \begin{ruledtabular}
		\begin{tabular}{cccccccc}
			$L$ & $m_1$ & $m_2$ & GoF & $ W_c$ & $\nu$ & $y$ & $g_c$ \\
			\hline
			\multirow{2}*{4-12} & 2 & 1 & 0.97 & 22.53 [22.50, 22.55] & 0.820 [0.710, 0.936] & 1.57 [1.48, 1.66] & 0.90 [0.88, 0.91] \\
			& 3 & 1 & 0.98 & 22.53 [22.51, 22.56] & 0.817 [0.722, 0.900]& 1.59 [1.50, 1.70] & 0.89 [0.88, 0.91] \\
			6-12 & 3 & 0 & 0.93 & 22.62 [22.58, 22.66] & 0.818 [0.713, 0.877] & 1.81 [1.55, 2.23] & 0.83 [0.81, 0.85]
		\end{tabular}
	\end{ruledtabular}
\end{table*}

\begin{figure*}
    \centering
    \includegraphics[width=0.45\textwidth]{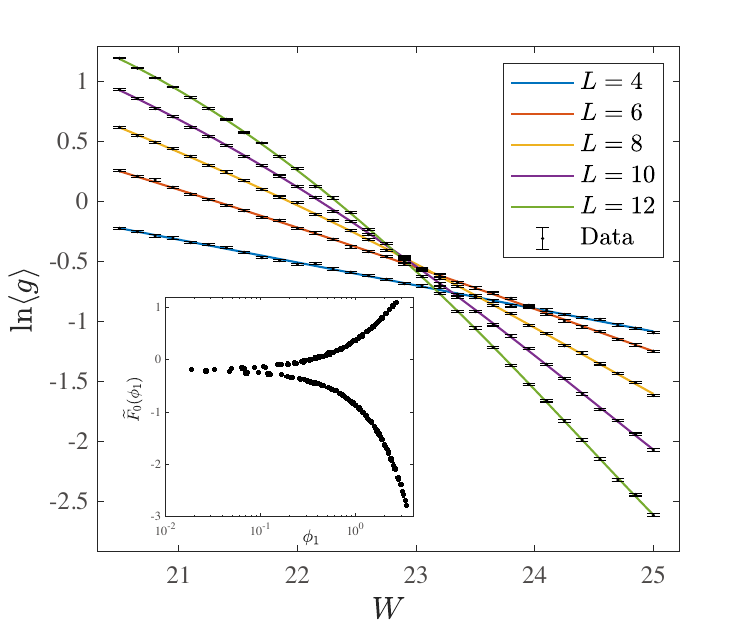}
    \qquad
    \includegraphics[width=0.45\textwidth]{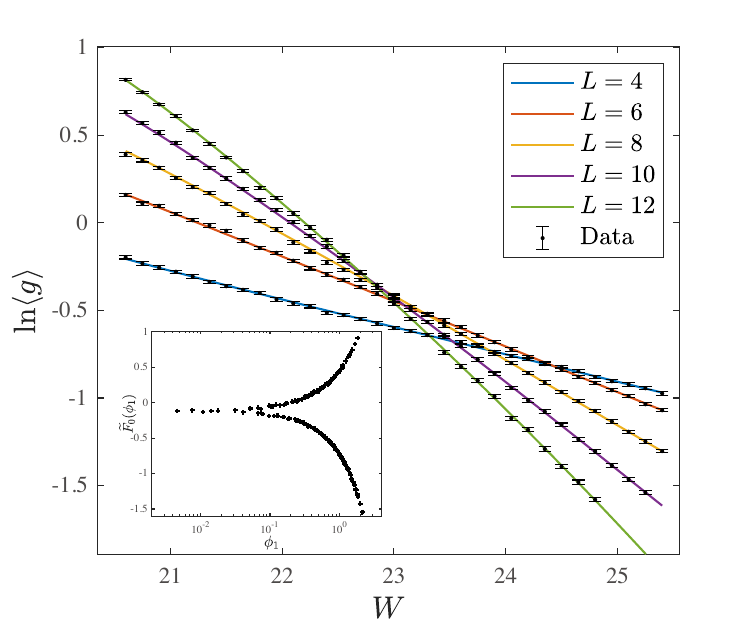}
    \caption{Dimensionless Landauer conductance as a function of disorder $W$ around the Anderson transition. The expansion order is $(n_1,n_2,m_1,m_2)=(3,1,2,0)$. \textbf{Left panel:} 4D symmetry class C. \textbf{Right panel:} 4D symmetry class CI. The colored solid lines are fitting functions and black dots with error bars are the numerical data.}
    \label{fig:4d}
\end{figure*}

\subsection{4D class C}
\label{subsec:4dC}
Symmetry class C describes disordered superconductors with spin-rotational symmetry but broken time-reversal symmetry. For this symmetry
class the spin quantum Hall effect occurs in two-dimensions \cite{SQHE1999}. We extend the 3D tight-binding model for class C of Ref. \cite{wang2021universality} to 4D,
\begin{align}
    \mathcal{H}_{\rm C} & =  \sum_{\mathbf{r},\mathbf{r'}}  c_{\mathbf{r}}^{\dagger} [H_{\rm C}]_{\mathbf{r}\mathbf{r}'} c_{\mathbf{r}'} \nonumber \\
    &= \sum_{\mathbf{r}} \left[ \sum_{\mu=1}^3 t c_{\mathbf{r}+\mathbf{e}_\mu}^{\dagger} c_{\mathbf{r}} +  t_{\parallel} c_{\mathbf{r}+\mathbf{e}_4}^{\dagger} c_{\mathbf{r}} \right. \nonumber \\
    &+ \left. \mathrm{i}t_{\perp} \bigg( c_{\mathbf{r}+\mathbf{e}_1}^{\dagger} \sigma_1 c_{\mathbf{r}} + \sum_{\mu=2,3} c_{\mathbf{r}+\mathbf{e}_{\mu}}^{\dagger} \sigma_2 c_{\mathbf{r}} \bigg) + \mathrm{H.c.} \right] \nonumber \\
    &+ \sum_{\mathbf{r}} (v_{\mathbf{r}}+\Delta) c_{\mathbf{r}}^{\dagger}\sigma_3 c_{\mathbf{r}}.
\end{align}
Here, $c_{\mathbf{r}}^{\dagger}$ is the creation operator on lattice site $\mathbf{r}=(x_1,x_2,x_3,x_4)$ where the two components act on spin, orbital or Nambu space depending on the nature of the system. The Hamiltonian has a particle-hole symmetry $U_P^{\dagger} H_{\rm C}^{*} U_P = - H_{\rm C}$ with
\begin{align}
	U_P = \delta_{\mathbf{r}\mathbf{r}'} e^{\mathrm{i}\pi\sum_{\mu=1}^4 \mathbf{r}\cdot \mathbf{e}_\mu} \sigma_2, \quad U_P^{T}=-U_P.
\end{align}
In the clean limit the Fourier transformation of the Hamiltonian is
\begin{align}
	h_{\rm C}(\bm{k}) &= 2 t_{\parallel} \cos k_4 + 2 t
 \sum_{\mu=1}^{3} \cos k_{\mu} + \Delta \sigma_3 \nonumber \\
	& - 2 t_{\perp} \left[ \sin k_1 \sigma_1 + \left( \sin k_1 + \sin k_2\right) \sigma_2 \right] .
\end{align}
For numerical simulations, we set $\Delta=0.5$, $t_{\perp}=t=1$ and $t_{\parallel}=0.8$ 
so that the clean system has a finite Fermi surface at $E_F=0$.
We calculate the two-terminal Landauer conductance $G$ using the transfer matrix method \cite{Pendry1992},
\begin{align}
    G = \frac{e^2}{h} g, \quad g=\mathrm{Tr} \left[ \tilde{t}^{\dagger}\tilde{t} \right],
\end{align}
where $\tilde{t}$ is the transmission matrix of the hypercubic samples of size $L^4$ along $w$ axis.
We impose periodic boundary conditions in directions the transverse to the current.
While the dimensionless conductance $g$ exhibits
fluctuations, various disorder average are well described by a scaling function like Eq. (\ref{eq:scaling}) \cite{ConductanceScaling1999,ConductanceScaling2003}.
We calculate $\ln \langle g \rangle$, and use the same nonlinear fitting procedures as described through Eq. (\ref{eq:scaling}-\ref{eq:GOF}). Each data point $\langle g \rangle$ is averaged over 5000–20000 samples to ensure a relative error smaller than 1\%.
The results for the critical exponent $\nu$ and other quantities are shown in Table \ref{tab:fit} (b). The fitting results are stable against change of  expansion order $m_1,m_2$ and the range of system size. 
Our estimate of the critical exponent for 4D class C is
\begin{align}
    \nu  = 0.72 \pm 0.02.
\end{align}
Note that the critical disorder $W_c$ and critical conductance $g_c$ are model-dependent, i.e.not universal.

\subsection{4D class CI}
Symmetry class CI describes disordered superconductors with both time-reversal symmetry and spin-rotational symmetry. Again, we extended the 3D class CI model of Ref. \cite{wang2021universality} to 4D
\begin{align}
    H_{\rm CI} &= \sum_{\mathbf{r},\mathbf{r'}}  c_{\mathbf{r}}^{\dagger} [H_{\rm CI}]_{\mathbf{r}\mathbf{r}'} c_{\mathbf{r}'} \nonumber \\
    &= \sum_{\mathbf{r}} \bigg[ \sum_{\mu=1}^3 t_{\perp} c_{\mathbf{r}+\mathbf{e}_\mu}^{\dagger} c_{\mathbf{r}} +  t_{\parallel} c_{\mathbf{r}+\mathbf{e}_4}^{\dagger} \sigma_3 c_{\mathbf{r}}  \nonumber \\
    &+ t'_{\parallel } c_{\mathbf{r}+\mathbf{e}_4}^{\dagger} \sigma_1 c_{\mathbf{r}} + \mathrm{H.c.} \bigg] + \sum_{\mathbf{r}} (v_{\mathbf{r}}+\Delta) c_{\mathbf{r}}^{\dagger}\sigma_1 c_{\mathbf{r}}.
\end{align}
The Hamiltonian is time-reversal symmetric since $H_{\rm CI}^* = H_{\rm CI}$, and has particle-hole symmetry $U_P^{\dagger} H_{\rm CI}^{*} U_P = - H_{\rm CI}$ given by
\begin{align}
    U_P = \delta_{\mathbf{r}\mathbf{r}'} e^{\mathrm{i}\pi \sum_{\mu=1}^3 \mathbf{r}\cdot\mathbf{e}_\mu} \sigma_2, \quad U_P^{T}=-U_P.
\end{align}
In the clean limit the Fourier transformation of the Hamiltonian is
\begin{align}
	h_{\rm CI}(\bm{k}) & = 2 t_{\perp} \sum_{\mu=1}^{3} \cos k_{\mu} + 2t_{\parallel}^\prime \cos k_4 \sigma_3 \nonumber \\
	& + (\Delta + 2 t_{\parallel} \cos k_4 ) \sigma_1 .
\end{align}
In numerical simulations of the two-terminal Landauer conductance, we chose $\Delta=1.2$, $t_{\perp}=1$ and $t_{\parallel} = t'_{\parallel} = 0.5$. 
Following the same procedures as described in the previous section, we estimate the critical exponent $\nu$ and other quantities.
The results are shown in Table \ref{tab:fit} (b).
Our estimate of the critical exponent $\nu$ for 4D class CI is
\begin{align}
   \nu = 0.83 \pm 0.04
\end{align}

\section{Comparison of Borel-Pad\'e predictions with numerical results}
\label{sec:comparison}
Referring to Table \ref{tab:A2}, we see that for classes C and CI in both 3D and 4D, the estimates of the
critical exponent obtained with the $[0,4]$ Borel-Pad\'e resummations are in good agreement with the numerical estimates.
For 3D class D the discrepancy is relatively large and even larger for 3D class DIII.
These are also the two symmetry classes where $d_l<2$ (see Table \ref{tab:dc1lcd}).
In addition we notice an inconsistency between our estimation of the critical exponent for 3D class DIII $\nu=0.96\pm 0.01$ and that in Ref. \cite{roy2017global} $\nu=0.85\pm 0.05$.
The model used in Ref. \cite{roy2017global} is essentially the same as here, but the data set of  Ref. \cite{roy2017global} is of smaller size and of lower numerical precision. However, we note the possibility that the weak topological indices may change the critical behavior of Anderson transition \cite{Xiao22}.

\begin{table}
\begin{ruledtabular}
    \centering
    \begin{tabular}{ccccc}
    \multicolumn{5}{l}{(a) 3D} \\
     & \multicolumn{2}{c}{Borel-Pad\'e with $A=2$} & \multicolumn{2}{c}{numerical} \\
     class & $[0,4]$ & $[1,3]$ & $\nu$ & Ref. \\
     \hline
     C & 1.056 & - & $0.996\pm 0.012$ & \cite{wang2021universality,Chalker09network} \\     
     CI & 1.107 & 1.822 & $1.17\pm0.02$ & \cite{wang2021universality} \\
     D & 0.823 & 0.858 & $0.87\pm0.03$ & \cite{wang2021universality}\\
     \multirow{2}*{DIII} & \multirow{2}*{0.751} & \multirow{2}*{0.674} & $0.85\pm 0.05$ & \cite{roy2017global} \\
      & & & $0.96\pm 0.01$ & * \\
     \\
     \multicolumn{5}{l}{(b) 4D} \\
     & \multicolumn{2}{c}{Borel-Pad\'e with $A=2$} & \multicolumn{2}{c}{numerical} \\
     class &  $[0,4]$ & $[1,3]$ & $\nu$ & Ref. \\
     \hline
     C & 0.714 & - & $0.70\pm0.02$ & * \\
     CI & 0.729 & 1.103 & $0.83\pm0.04$ & * \\
     D & 0.640 & 0.666 & - & -\\
     DIII & 0.616 & 0.589 & - & -\\
    \end{tabular}
    \caption{\label{tab:A2}Critical exponents $\nu$ of the BdG symmetry classes in 3D and 4D obtained from order $[m,n]$ Borel-Pad\'e resummation of the $\beta$-function with $A=2$, and numerical simulations.
    Here, $*$ indicates the numerical estimates in this paper, whereas $-$ indicates that the value is yet to be determined.
We omit the $[1,3]$ resummation with $A=2$ for class C because the resummed $\beta$-function is not monotonic and has two unphysical fixed points.  This is also the case for $A=1$.}
\end{ruledtabular}
\end{table}

We have resummed the series for the $\beta$-function in such a way that Eq. (\ref{eq:lim_tdfdt}) is satisfied.
This resummation means that in the localised regime the $\beta$-function will behave like $A\ln g$ up to a constant.
It would then seem more natural to set $A=1$ rather than $A=2$.
However, the former choice does not yield the correct limiting behavior Eq. (\ref{eq:nu_limit_half}).
For reference, we also tabulate the estimates of the critical exponents calculated 
from the truncated $\beta$-function series without resummation and 
from the Borel-Pad\'e analysis with $A=1$ in Table \ref{tab:A1}.
Without resummation, we obtain estimates that violate the Chayes inequality $\nu\ge 2/d$ \cite{chayes1986finite}.
With $A=1$, the estimates satisfy the Chayes inequality but are in poorer agreement with the numerical estimates compared
with $A=2$.

\begin{table}
\begin{ruledtabular}
    \centering
    \begin{tabular}{cccc}
    \multicolumn{4}{l}{(a) 3D} \\
     & no resummation & \multicolumn{2}{c}{Borel-Pad\'e with $A=1$}  \\
     class & & $[0,4]$ & $[1,3]$  \\
     \hline
     C & 0.471 & 1.446 & -  \\
     CI & 0.555 & 1.478 & 2.131 \\
     D & 0.187 & 1.254 & 1.249 \\
     DIII & 0.151 & 1.202 & 1.088 \\
     \\
     \multicolumn{4}{l}{(b) 4D} \\
     & no resummation & \multicolumn{2}{c}{Borel-Pad\'e with $A=1$}  \\
     class &  & $[0,4]$ & $[1,3]$   \\
     \hline
     C & 0.200 & 1.122 & -  \\
     CI & 0.217 & 1.129 & 1.602  \\
     D & 0.103 & 1.075 & 1.079  \\
     DIII & 0.091 & 1.062 & 1.026  \\
    \end{tabular}
    \caption{\label{tab:A1}Critical exponents $\nu$ of the BdG symmetry classes in 3D and 4D obtained from $\beta$-function series without resummation and order $[m,n]$ Borel-Pad\'e resummation with $A=1$.}
\end{ruledtabular}
\end{table}

\section{Summary and Discussion}
\label{sec:summary}

In this paper, we have studied the Anderson transition in the BdG symmetry classes both analytically and numerically.
We applied the Borel-Pad\'e resummation method to the known perturbative results for the
NL$\sigma$M to estimate the critical exponents in 3D and 4D.
We also reported numerical simulations of class DIII in 3D, and classes C and CI
in 4D, and compared the results of the resummation method with the results of the resummations and previously published
work.
We find that the results of the Borel–Pad\'e analysis provide estimates of the critical exponent with the numerical
estimates provided the limiting behaviour Eq. (\ref{eq:nu_limit_half}) is imposed during the resummation.
In principal, the NL$\sigma$M theory of Anderson localization and its renormalization analysis 
in $d=2+\epsilon$ dimensions are valid only when $\epsilon$ is small, i.e., the Anderson transition
occurs under weak disorder.
Nonetheless, our results show that the perturbative $\beta$-functions can provide useful information concerning
critical properties in 3D and 4D.

The estimations of the critical exponents in BdG symmetry classes based on the Borel-Pad\'e resummation methods with the assumption of infinite upper critical dimension match the numerical results better.
This suggest that the upper critical critical dimension $d_u$ may be infinite for the Anderson localization in BdG symmetry classes.
Previous theoretical works have argued that in noncompact NL$\sigma$M, the upper critical dimension is infinite\cite{codello2009fixed,efremov2021nonlinear}, 
which seems to be consistent with the numerical results and estimation of Borel-Pad\'e resummation method in this work.
Further theoretical efforts are needed to conform these observations.

Recently, it has been pointed out that the NL$\sigma$M model characterizes the measurement-induced phase transition in quantum circuits\cite{Fava23}. This scenario involves a replica number $N$ equal to 1.
The resummation method discussed in this paper is also applicable to that case, allowing for the prediction of critical exponents in quantum circuit systems.

{\bf Acknowledgments}
We thank and Ryuichi Shindou, Ferdinand Evers and Alexander D. Mirlin for fruitful discussions.
T.W. was supported by the National Basic Research Programs of China (Grant No. 2019YFA0308401) and the National Natural Science Foundation of China (Grants No. 11674011 and No. 12074008). Z.P. was supported by National Natural Science Foundation of China (No. 12147104). T.O. and K.S. were supported by JSPS KAKENHI Grants 19H00658, and T.O. was supported by JSPS KAKENHI 22H05114.

\bibliography{References}

\end{document}